# Simulation of Skin Stretching around the Forehead Wrinkles in Rhytidectomy

**Ping Zhou[a,*], Shuo Huang[a,b], Qiang Chen[a], Siyuan He[a], Guochao Cai[a]**

[a]School of Biological Sciences & Medical Engineering, Southeast University, 210096 Nanjing, P.R. China
[b]Shanghai United-imaging Healthcare Co., Ltd, 201807 Shanghai, P. R. China

**Abstract. Objective:** Skin stretching around the forehead wrinkles is an important method in rhytidectomy. Proper parameters are required to evaluate the surgical effect. In this paper, a simulation method was proposed to obtain the parameters. **Methods:** Three-dimensional point cloud data with a resolution of 50 μm were employed. First, a smooth supporting contour under the wrinkled forehead was generated via b-spline interpolation and extrapolation to constrain the deformation of the wrinkled zone. Then, based on the vector formed intrinsic finite element (VFIFE) algorithm, the simulation was implemented in Matlab for the deformation of wrinkled forehead skin in the stretching process. Finally, the stress distribution and the residual wrinkles of forehead skin were employed to evaluate the surgical effect. **Results:** Although the residual wrinkles are similar when forehead wrinkles are finitely stretched, their stress distribution changes greatly. This indicates that the stress distribution in the skin is effective to evaluate the surgical effect, and the forehead wrinkles are easily to be overstretched, which may lead to potential skin injuries. **Conclusion:** The simulation method can predict stress distribution and residual wrinkles after forehead wrinkle stretching surgery, which can be potentially used to control the surgical process and further reduce risks of skin injury.

*Ping Zhou, E-mail: zhouping@seu.edu.cn

## 1. INTRODUCTION

Wrinkle is a direct clue indicating a person's age, thus, wrinkle elimination is an important issue of the beauty industry[1]. Rhytidectomy, as an effective way to eliminate facial wrinkles, is a surgery to smooth wrinkles which cannot be removed by drugs and other physical methods[2].

At present, one of clinical strategies to eliminate the forehead wrinkle is known as coronal-incisional method[2,3]. The coronal-incision method is incising the forehead skin at the hairline and then stretching the incision to achieve the effect of wrinkle elimination. It is well-accepted that the method is reliable, but the coronal-incision rhytidectomy does cause complications such as scar and alopecia, as the skin is easily to be overstretched. In other words, overstretching the forehead skins will injure skin fibres, and thus leads to the skin injury. Meanwhile, the main loaded area studied is influenced by the boundary conditions constrained by its surrounding tissue, which will cause severe stress concentration at these boundaries. Therefore, it is crucial to use appropriate indexes to evaluate the surgical effect before the operations.

However, it is difficult to evaluate the biomechanical effect during the surgical process, thus the surgical injury is difficult to be measured and avoided[3]. In this regard, the numerical method provides an effective way to solve the biomechanical issues. For example, basing on the Newton's Second Laws, Ting et al (2004) proposed a simple vector formed intrinsic finite element method (VFIFE), which considered the nonlinearity without dealing with the structural stiffness matrix, and the method was proved to be precise and effective in processing large structural deformation[4-9].

In this paper, the computer simulation method VFIFE was employed to obtain the residual wrinkles (the visual performance) and stress distribution in the forehead skin after surgery, which were used to evaluate the surgical effect of the forehead wrinkles elimination before the stretching operation. To the best of authors' knowledge, this is the first work to simulate the stretching process of the forehead wrinkle elimination.

## 2. METHODS

*2.1 Acquisition of point cloud data*

A three-dimensional measurement system in the experiments applies Phase Measurement Profilometry method and multifrequency heterodyne principle to acquire forehead wrinkles data[10,11]. The system is equipped with a BENQ GP1 projector and two MVD 120SC CCD cameras. Point cloud data of forehead wrinkles acquired with this system has a high resolution of 50 μm, which is able to present the wrinkles clearly[10,11]. Then, three-dimensional point cloud data of forehead wrinkles is input into the simulation, in particular, a set of point cloud data of selected forehead wrinkles is shown in Fig. 1.

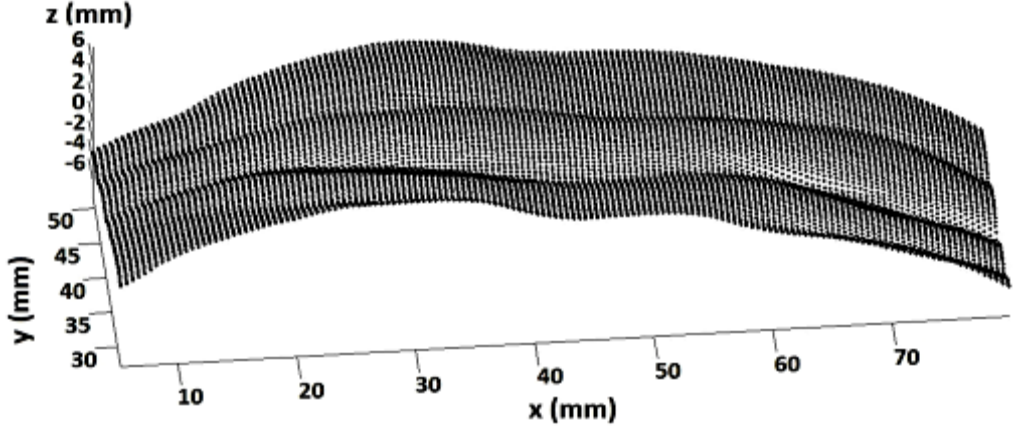

**Fig. 1** A set of point cloud data of the selected forehead wrinkles.

*2.2 Extraction of the supporting contour*

Since the tissues (muscle and bone) under the stretched forehead skin restrain its deformation in the stretching process, it is very important to define a three-dimensional rigid supporting contour **Z**$_{sup}$, which restrains the movement of forehead skin in the simulation. A method to extract **Z**$_{sup}$ is described below.

As the wrinkle is a texture which appears gradually on human's face during growth, the forehead wrinkles **Z**$_c$ shown in Fig. 1 are here regarded as the superposition of a smooth forehead contour **Z**$_s$ and wrinkles $\varepsilon$, i.e.:

$$\mathbf{Z_c} = \mathbf{Z_s} + \varepsilon \tag{1}$$

In the no-wrinkle case, $\varepsilon=0$, and **Z**$_c$=**Z**$_s$. To obtain **Z**$_s$, a fitting surface $Z_f(x,y)$ to **Z**$_c$ is firstly calculated by implementing a spline function as:

$$Z_f(x,y) = \sum_{j=1}^{n_x \times n_y} C_j N_{j,p}(x,y) \tag{2}$$

where $C_j$ is the control points, $N_{j,p}$ is the fitting polynomial, the subscript j ranges from 1 to $n_x \times n_y$ ($n_x$ and $n_y$ are the numbers of nodes along the x axis and y axis, respectively), and the subscript p is the order of the fitting polynomial, which affects the smoothness of surface and is here set to be 3. Alternatively, the fitting function (2) can be re-expressed as a linear algebra equation:

$$\mathbf{Z_f} = \mathbf{NC} \tag{3}$$

where the control point matrix $\mathbf{C}$ is of length $n_x \times n_y$, thus the matrix $\mathbf{N}$ of the fitting polynomial has $n_x \times n_y$ rows and $n_x \times n_y$ columns corresponding to each data point in the forehead wrinkles. To make $\mathbf{Z_f}$ continuous at a point, the value of the first partial derivatives at its neighboring points is forced to be equal, which results in a second partial derivative of linear equations in the form of

$$\mathbf{BC} = \mathbf{0} \tag{4}$$

where $\mathbf{B}$ is the approximate value of the second partial derivatives by using finite difference of the fitting surface $\mathbf{Z_f}$ at neighboring points of the point. The sum of the second partial derivatives is forced to be zero. To obtain the matrix $\mathbf{C}$, Eq. (3) and Eq. (4) are solved in combination:

$$\|(\mathbf{NC} - \mathbf{Z_{i-1}})\|^2 + \lambda \|\mathbf{BC}\|^2 \qquad i \geq 1, \quad \mathbf{Z_0} = \mathbf{Z_c} \tag{5}$$

where $\lambda$ is the weight of the sum of the second partial derivatives, $\mathbf{Z_{i-1}}$ is the result of the (i-1)th iteration. The matrix $\mathbf{C}$ under a given $\lambda$ can be solved through the calculation of function argmin, and then $\mathbf{Z_f}$ can be obtained[12]. With the surfaces $\mathbf{Z_c}$ and $\mathbf{Z_f}$, the following two steps are implemented to calculate $\mathbf{Z_s}$ as:

(1) The difference in the z axis between the corresponding points on $\mathbf{Z_c}$ and $\mathbf{Z_f}$ is compared to a threshold (here, the threshold is set to be 0.15 mm) to eliminate the points in wrinkled areas on the surface $\mathbf{Z_c}$, and then a surface $\mathbf{Z_d}$ that do not contain wrinkles is generated.

(2) $Z_s$ is obtained by expanding $Z_d$ with the combination of b-spline interpolation and extrapolation methods[13], see Fig. 2(a).

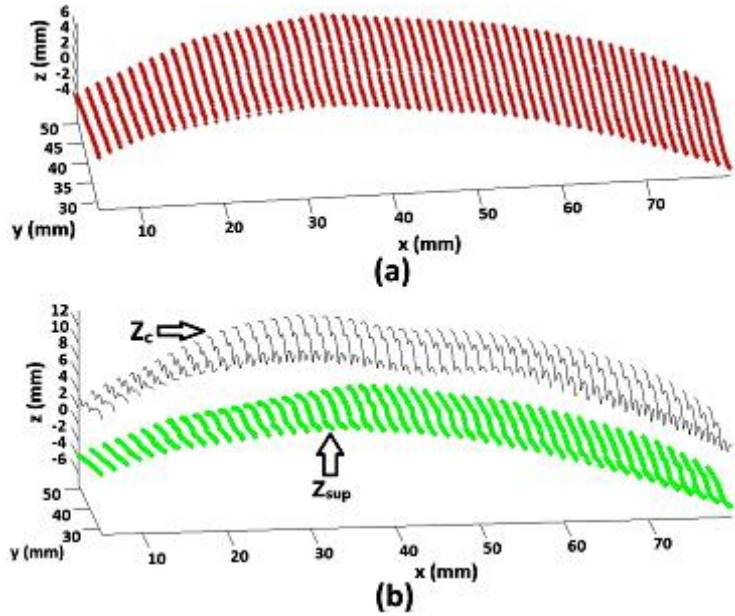

**Fig. 2** The extraction of supporting contour. (a) The smooth forehead contour $Z_s$. (b) Supporting contour $Z_{sup}$ and the forehead wrinkles $Z_c$.

Finally, $Z_{sup}$ is obtained by subtracting the maximal difference between $Z_s$ and $Z_c$ from $Z_s$, see Fig. 2b, where $Z_c$ is moved by 6 mm in the positive z direction for the sake of clarity. The flowchart of the supporting contour extraction $Z_{sup}$ is shown in Fig. 3.

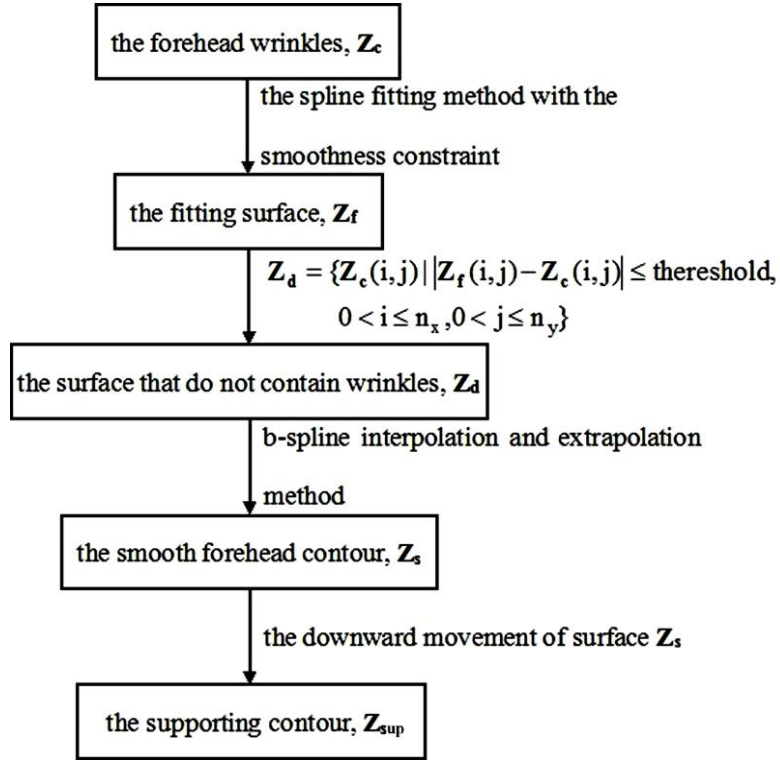

**Fig. 3** The flowchart of supporting contour extraction.

*2.3 Forehead Wrinkle Stretching Method*

*2.3.1 The Finite Element Meshing*

Vector form intrinsic finite element (VFIFE) method[4-9] is used to simulate the stretching process of the forehead wrinkles. The VFIFE method first discretizes a structure into particles with units connecting the particles, and then calculates the structural shape, location, and stress distribution by the displacements of particles.

Here, the forehead wrinkles are discretized by the method, and the discrete particles have a same row distance and column distance, thus, the distribution of skin mass on the particles is approximately uniform[4,10,11], and the structure unit only carries tensile load. To simulate the real surgical conditions, the bottom and lateral sides of the forehead skin model are fixed, and the skin model is stretched by uniformly tensioning the top side with a constant speed. The square element type is utilized, and the meshed model of the forehead wrinkles is shown in Fig. 4.

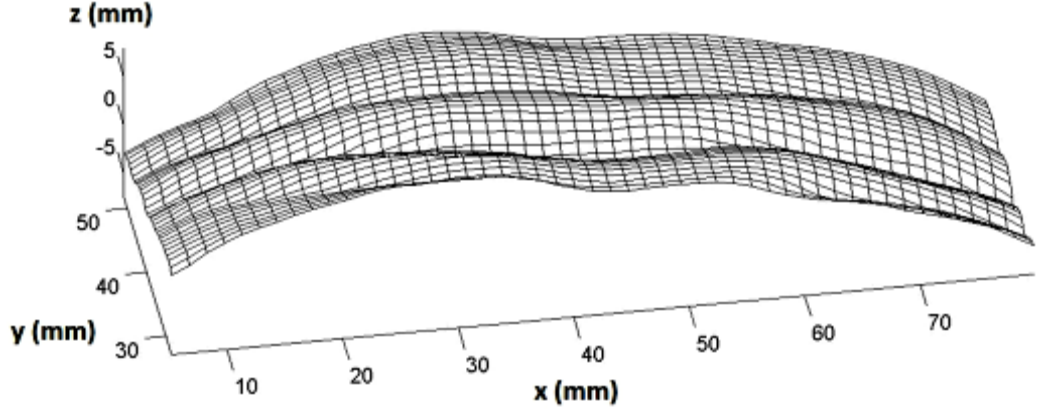

**Fig. 4** The finite element meshing of the forehead wrinkles.

Moreover, since the stretch in the wrinkle elimination surgeries is limited, and the skin is under small deformation[14-18], the small deformation theory under which the skin stress-strain relationship obeys Hooke's Law is considered.

*2.3.2 Force calculation of particle*

As statement of the VFIFE method, to calculate the stress distribution of the forehead wrinkle, the displacement $\mathbf{x_P}$ of an arbitrary particle P of the forehead wrinkle is needed. According to the Newton's second law, the force of acting on the particle P can be solved as:

$$m_P \frac{d^2 \mathbf{x_P}}{dt^2} = \mathbf{F}_{P,t} \tag{6a}$$

with

$$\mathbf{F}_{P,t} = \mathbf{f}_{P,t}^{int} + \mathbf{f}_{P,t}^{ext} \tag{6b}$$

$$t = t_0 + n(\Delta t) \quad n \in N \tag{6c}$$

where $m_P$ is the mass of the particle, $\mathbf{F}_{P,t}$ is the tensile force vector applied on the particle P by the surrounding tensile structure units at time t, $\mathbf{f}_{P,t}^{int}$ and $\mathbf{f}_{P,t}^{ext}$ are the internal force and external force, respectively, $t_0 = 0$ is the initial time, n is the number of the time step.

$\mathbf{f}_{P,t}^{int}$ is the resultant of the forces of the tensile structures surrounding particle P at time t, and according to the discrete particle motion model in Fig. 5, it is the sum of $\mathbf{f}_{P,t}^{i}$, which is the internal force exerted by the ith (i = 1, 2, 3, 4) surrounding structure unit at time t. As the stress-strain relationship of the structure units obeys Hook's law in the case of small deformation, thus we have:

$$\mathbf{f}_{P,t}^{i} = \mathbf{f}_{P,0}^{i} + \frac{ES}{\|\mathbf{U}_{i,0}\|} \Delta \mathbf{U}_{i,t}, \; i = 1,2,3,4. \tag{7}$$

where $\mathbf{f}_{P,0}^{i}$ and $\mathbf{f}_{P,t}^{i}$ are internal forces of the ith surrounding structure unit at the time 0 and t, respectively, E is the elastic modulus of skin, S is the projecting area of the square element on the plane perpendicular to the loading direction, $\mathbf{U}_{i,0}$ and $\Delta \mathbf{U}_{i,t} = \mathbf{U}_{i,t} - \mathbf{U}_{i,0}$ are the initial length and the elongation vectors of the i-th tensile structure unit at the time t, respectively. The cross-sectional area in the same direction is approximately equal, as the elongation of skin is not very high during the simulation[2]. At the initial time 0, particles in the forehead wrinkles are unloaded, thus, $\mathbf{f}_{P,0}^{i}$ disappears, and Eq. (7) is rewritten as:

$$\mathbf{f}_{P,t}^{i} = \frac{ES}{\|\mathbf{U}_{i,0}\|} \Delta \mathbf{U}_{i,t}, \; i = 1,2,3,4. \tag{8}$$

As stated in Section 2.3.1, skin only carries tensile loads, so that the internal force is set to be zero when the internal force of the structure units is negative, *i.e.*,

$$\mathbf{f}_{P,t}^{i} = \{\mathbf{0} \mid \|\mathbf{U}_{i,t}\| - \|\mathbf{U}_{i,0}\| \leq 0, \; i = 1,2,3,4\} \tag{9}$$

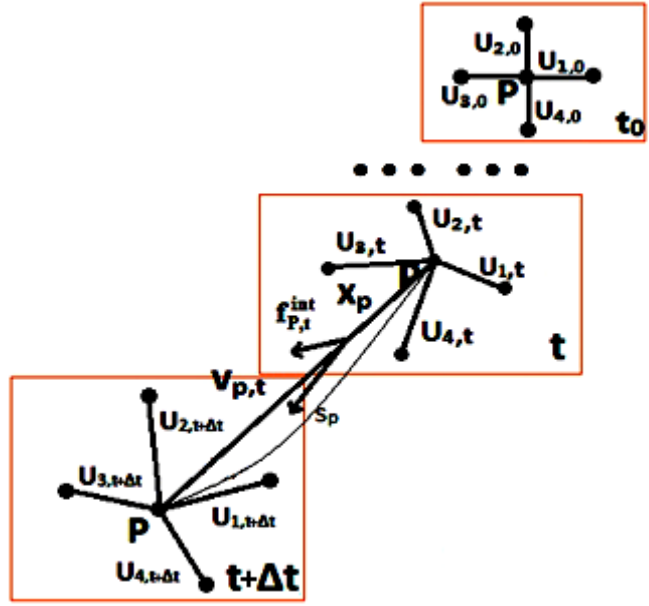

**Fig. 5** Discrete particle motion model, $s_P$ is the trajectory of the particle P.

Here, $\mathbf{f}_{P,t}^{ext}$ includes two parts. One is the tensile load exerted on the particles of the incised boundary. To facilitate the simulation, the particles of the boundary are stretched with a constant speed, so that these particles are approximately in a quasi-static equilibrium. The other is the interaction between the skin and supporting contour $\mathbf{Z_{sup}}$, and the gravity of particles is ignored because it is much smaller than the tensile load. The interaction is described by a collision model shown in Fig. 6. In this model, the supporting contour $\mathbf{Z_{sup}}$ is meshed in the same way as forehead wrinkles' mesh in Fig. 4, and Each element of the supporting contour $\mathbf{Z_{sup}}$ is regarded as a tiny plane.

From Fig. 6, at time t, the acceleration vector of the particle P at the position $\mathbf{b}_t$ is $\mathbf{a}_t$ and its speed vector is $\mathbf{v}_t$. If the supporting contour does not exist, the particle P at the position $\mathbf{b}_t$ will move to $\mathbf{b}_{t+\Delta t}^{virt}$, and correspondingly, its speed vector changes to $\mathbf{v}_{t+\Delta t}^{virt}$. On the contrary, if the supporting contour is considered, the particle P collides with the supporting contour at the point O, thus the position of the particle P shifts from $\mathbf{b}_{t+\Delta t}^{virt}$ to $\mathbf{b}_{t+\Delta t}$, and its speed vector changes to $\mathbf{v}_{t+\Delta t}$. We here select an arbitrary point Q on the supporting contour, which is used to judge whether the point P collides with the supporting contour. $\mathbf{w_z}$ is the distance from the point Q to the position $\mathbf{b}_{t+\Delta t}^{virt}$. Meanwhile, $\mathbf{u}$ is a unit normal vector of the point O on the supporting surface.

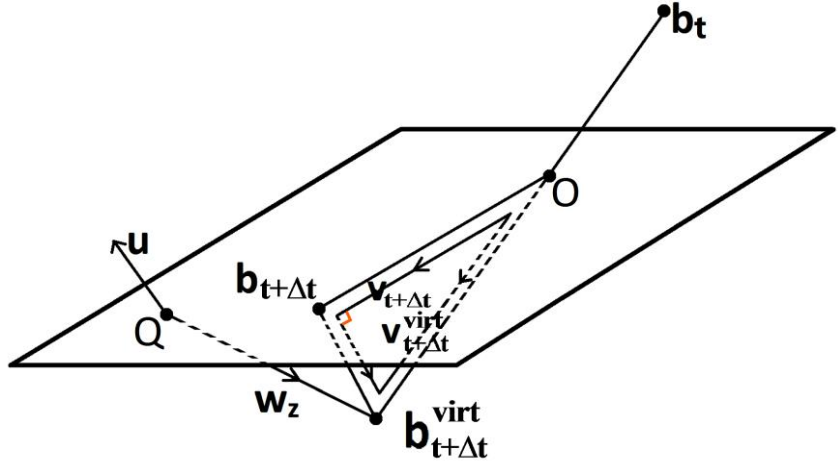

**Fig. 6** The collision model.

Moreover, according to the geometrical relationship among the vectors in Fig. 6, when $\mathbf{u} \cdot \mathbf{w}_\mathbf{z}^\mathbf{T} < 0$, the particle P collides with the supporting contour at time t+Δt. It is noticed that the velocity component perpendicular to the supporting contour becomes zero, since the collision between the skin and the supporting contour is perfectly inelastic. Therefore, the point P slides on the supporting contour with speed $\mathbf{v}_{t+\Delta t}$ after the collision from point O to its final position $\mathbf{b}_{t+\Delta t}$. The speed and position of the particle P at time t+Δt are derived as:

$$\mathbf{v}_{t+\Delta t} = \mathbf{v}_{t+\Delta t}^{virt} - \left( \mathbf{v}_{t+\Delta t}^{virt} \cdot \mathbf{u}^\mathbf{T} \right) \cdot \mathbf{u} \tag{10}$$

$$\mathbf{b}_{t+\Delta t} = \mathbf{b}_{t+\Delta t}^{virt} - \left( \mathbf{w}_\mathbf{z} \cdot \mathbf{u}^\mathbf{T} \right) \cdot \mathbf{u} \tag{11}$$

In order to obtain the equilibrium solution of Eq. (10) and Eq. (11), the loading progress maintains a low constant speed (0.1 mm/s) to reduce the vibration amplitude of the structure units. Moreover, an energy dissipation method is employed to make the structure units reach the quasi-static equilibrium eventually[9].

Besides, considering factors during the movement of the forehead skin on the supporting surface $\mathbf{Z_{sup}}$, such as friction, which may influence the particles' equilibrium, a damping force $\mathbf{f_{ad}}$ is added to Eq. (6a), that is,

$$m_P \frac{d^2 \mathbf{x_P}}{dt^2} = \mathbf{F_P} + \mathbf{f_{ad}} \tag{12}$$

with

$$\mathbf{f_{ad}} = -\alpha m_P \frac{d\mathbf{x_P}}{dt} \tag{13}$$

where α denotes the damping coefficient. Again, according to Newton's second law, an iterative method with equal time step Δt is employed to calculate the speed and displacement of the particle P as:

$$\begin{aligned}
\mathbf{x_{P,1}} &= \mathbf{x_{P,0}} + \mathbf{v_{P,0}} \times (\Delta t) \\
\mathbf{x_{P,n+1}} &= \mathbf{x_{P,n}} + \mathbf{v_{P,n}} \times (\Delta t) + \frac{1}{2} \times \frac{\mathbf{F_{P,n}} + \mathbf{f_{ad,n}}}{m_P} \times (\Delta t)^2, n \geq 1 \\
\mathbf{v_{P,n+1}} &= \mathbf{v_{P,n}} + \frac{\mathbf{F_{P,n}} + \mathbf{f_{ad,n}}}{m_P} \times (\Delta t), n \geq 0
\end{aligned} \tag{14}$$

where $\mathbf{x_{P,0}}$ and $\mathbf{v_{P,0}}$ denote the initial position and speed of the particle P at time 0, respectively.

Finally, combining Eqs. (12-14), forehead wrinkles reach the equilibrium state as the total kinetic energy decreases below a threshold[9]. Usually the threshold is $10^{-12}$ J, which means that the mean speed of forehead wrinkles is lower than 0.01 mm/s.

The flowchart of the forehead wrinkle stretching method is shown in Fig. 7.

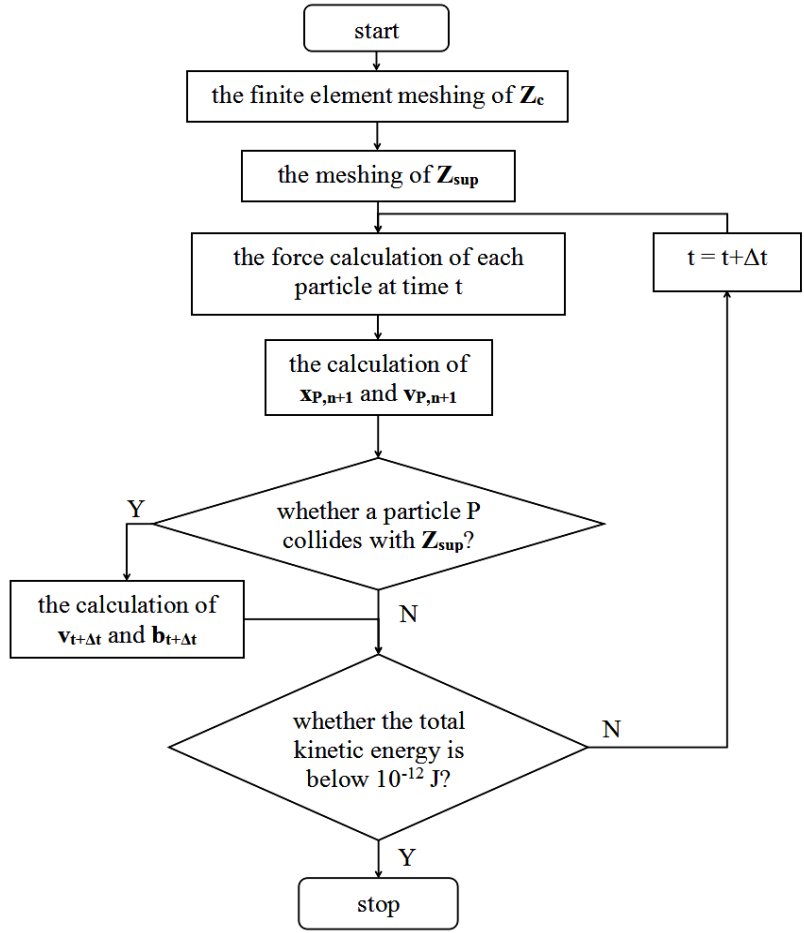

**Fig. 7** The flowchart of the forehead wrinkle stretching method.

## 3. PARAMETRIC EVALUATION

As mentioned before, inappropriate surgical parameters may cause the stress concentration at local zones, and further injures the skin. Therefore, it is of great importance to evaluate the simulation results by appropriate surgical parameters, and here the stress distribution and the residual forehead wrinkles are analyzed.

*3.1 Stress distribution*

According to the stress-strain relationship, when the stress in the *x* direction is more than $2.5 \times 10^5$ Pa, or the stress in the y direction is more than $5.0 \times 10^5$ Pa, skin fiber is considered to suffer local injury[15]. The stress distribution is measured by the Cauchy stresses of each element in the *x* and *y* directions, and employing the force in equation (8), the stresses are calculated as:

$$\sigma_y = \frac{f_y}{A_x} = \frac{\frac{EA_x}{U_{y,0}}\Delta U_{y,t}}{A_x} = \frac{E\Delta U_{y,t}}{U_{y,0}}$$

$$\sigma_x = \frac{f_x}{A_y} = \frac{\frac{EA_y}{U_{x,0}}\Delta U_{x,t}}{A_y} = \frac{E\Delta U_{x,t}}{U_{x,0}}$$

(15)

where $A_x$ and $A_y$ denote the cross-sectional areas in the x and y directions, respectively, $U_{x,0}$ and $U_{y,0}$ are the initial length of skin element in the x and y directions, respectively, and $\Delta U_{x,t}$ and $\Delta U_{y,t}$ denote the elongation of the skin element after being stretched at time t, respectively.

*3.2 The residual forehead wrinkles*

Here the parameters of the residual forehead wrinkles include the number of wrinkles, average depth, maximum depth and wrinkled area[19-22], besides, three new parameters, P60, P80 and Am, are also proposed to evaluate the difference between the simulation results under different loading cases. P60 and P80 stands for the area fractions of wrinkles, whose depth is over 60% and 80% of the maximum depth of the initially unstretched forehead wrinkles, respectively. Am is the area of wrinkles whose depth is greater than the mean depth of the initially unstretched forehead wrinkles.

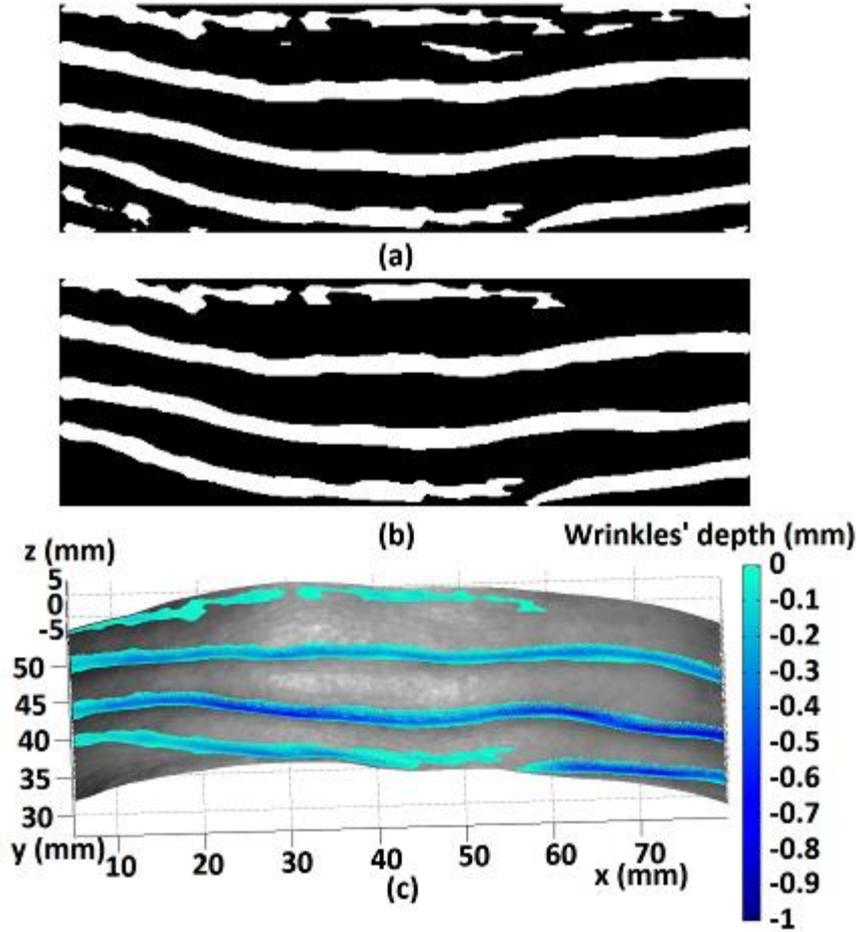

**Fig. 8** The extraction of the surface ε with wrinkles. (a) binary image of wrinkles after preliminary extraction. (b) binary image of the noise-removed surface. (c) the final three dimensional extraction results.

To obtain the above evaluation parameters, the residual wrinkles after being stretched need to be extracted. Like the method extracting the surface $Z_d$ containing no wrinkles in Section 2.2, the surface ε with wrinkles, are acquired by extracting points in the forehead wrinkles $Z_c$ using the distance in z direction between a point on $Z_c$ and the corresponding point on the fitting surface $Z_f$. If the distance of a point is larger than a threshold, the point is included in the preliminary extraction result. Indeed, there exists noisy points in the preliminary extraction result, as shown in Fig. 8a, where for the sake of clarity, wrinkles are projected to xy plane, and the white strips are the preliminary extracted areas. Then, the noise-removed surface was obtained by using the morphology method proposed in our previous work[23], see Fig. 8b, and the final three-dimensional extraction results are shown in Fig. 8c, where wrinkles are displayed in a color map which reflects their original depth in Fig. 1.

## 4. RESULTS AND DISCUSSIONS

Evaluation is performed on five volunteers' datasets aging from 20 to 40 years old in this part. In order to improve surgical effect without injuring skin, two parameters will be optimized: one is the loading displacement $L_d$, and the other is the distance $d_{fs}$ between stretching line and the fixed sides, see Fig. 9.

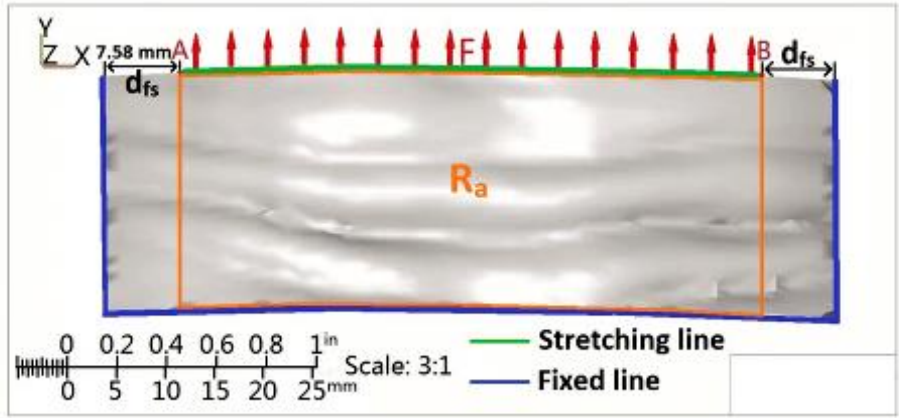

**Fig. 9** Diagram of forehead wrinkles' stretching.

*4.1 Optimizing $L_d$*

To optimize the loading displacement $L_d$, three cases with $L_d$ = 2 mm, 3 mm, 4 mm and constant $d_{fs}$ = 7.58 mm are studied, and the result of the depth of the residual wrinkles are plotted in Fig. 10. We can see that compared to the case of $L_d$ = 2 mm, there is no apparent residual wrinkles except the regions close to the fixed lateral sides when $L_d$ is 3 mm and 4 mm. This indicates that the skin should be stretched more than 3 mm to obtain a better surgical performance.

Moreover, the surgical effects are thoroughly evaluated by the parameters in Section 3.2, which are listed in Table 1, and the first row in Table 1 is the results under no load. Table 1 shows that both wrinkled area and Am decreases sharply when $L_d$ changes from 0 mm to 3 mm. Compared to the case of $L_d$ = 3mm, there are no obvious improvement in both area and Am for the case of $L_d$ = 4 mm. Moreover, when $L_d$ varies from 0 mm to 2 mm, Max depth and Mean depth decrease, while they increase when $L_d$ from 2mm to 4 mm. Especially, for the three loading cases, they reach their maximums when $L_d$ = 4 mm, and this indicates that the forehead wrinkles are deepened because of the constraint from fixed lateral sides.

For Fig. 10, the stress distribution in the x and y directions are plotted in Fig. 11 and Fig. 12, respectively. From Fig. 11, the stress in the x direction increases with the increase of loading displacement, and their stress distribution resembles wrinkles' distribution. In particular, according to the critical value of the skin injury ($2.5 \times 10^5$ Pa)[15], the forehead wrinkles are not overstretched. Moreover, the largest stress in Fig. 11c marked by arrows is two times of that in Fig. 11b. This is because the two ends of the top side (i.e. incision) are fixed, the deformation is constrained. From Fig. 12, increasing the loading displacement, the stress in the y direction also increases, and forehead wrinkles are overstretched when $L_d$ = 4 mm (Fig. 12c), as the maximum stress exceeds $5.0 \times 10^5$ Pa[15], beyond which the skin is injured. The largest stress marked by arrows in Fig. 12c is almost 2.5 times larger than that in Fig. 12b. It is seen that the stress in the y direction is band-like distributed, and this is because due to the Poisson's effect, the wrinkles is compressed in the x direction as the loading displacement is applied in the y direction.

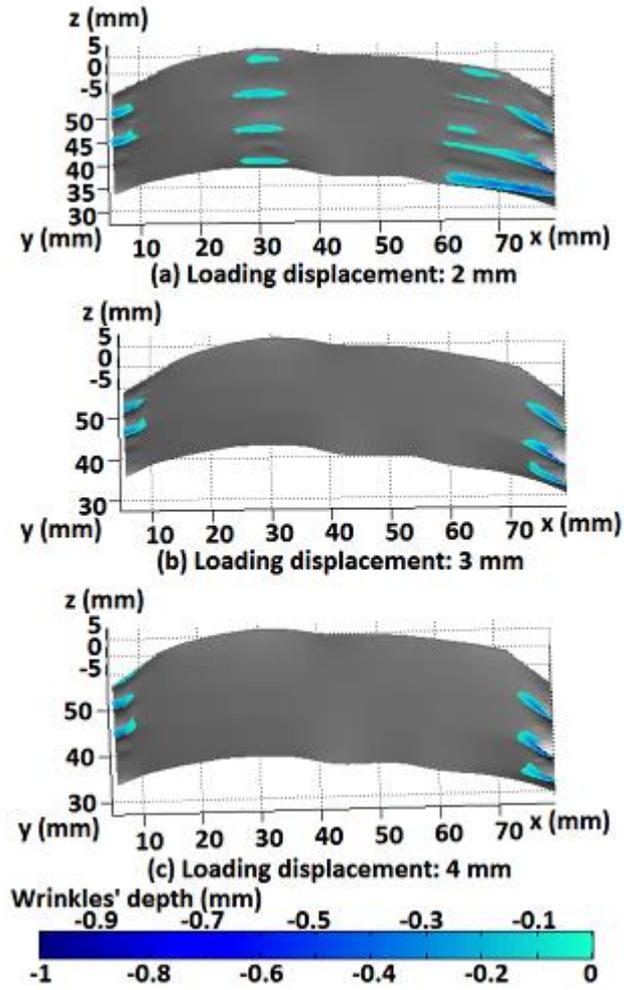

**Fig. 10** The stretching surgery simulations of forehead wrinkles in Fig. 1 under a constant $d_{fs}$ of 7.58 mm, and different $L_d$ of 2 mm, 3 mm and 4 mm.

**Table 1** The Quantified Parameters of the Residual Wrinkles of Simulation Results in Fig. 10

| No. | $L_d$/mm | $d_{fs}$/mm | Number of wrinkles | Max depth /mm | Mean depth /mm | P60 | P80 | Am/mm² | Area/mm² |
|---|---|---|---|---|---|---|---|---|---|
| 0 | 0 | 0 | 6 | 0.77 | 0.25 | 0.1132 | 0.0370 | 165.0290 | 390.6769 |
| 1 | 2 | 7.58 | 12 | 0.74 | 0.16 | 0.0148 | 0.0040 | 30.4162 | 145.6867 |
| 2 | 3 | 7.58 | 5 | 0.75 | 0.19 | 0.0466 | 0.0175 | 14.1482 | 53.6958 |
| 3 | 4 | 7.58 | 6 | 0.76 | 0.19 | 0.0486 | 0.0196 | 11.8868 | 50.9789 |

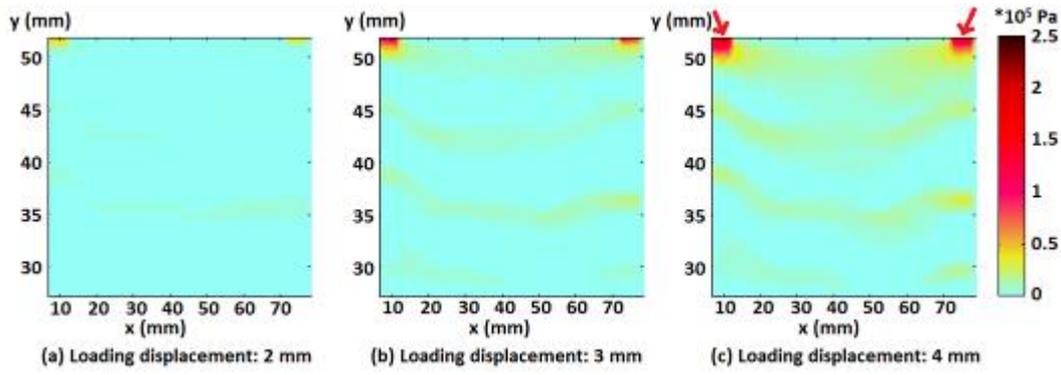

**Fig. 11** Stress distribution maps in the x direction of simulation results in Fig. 10.

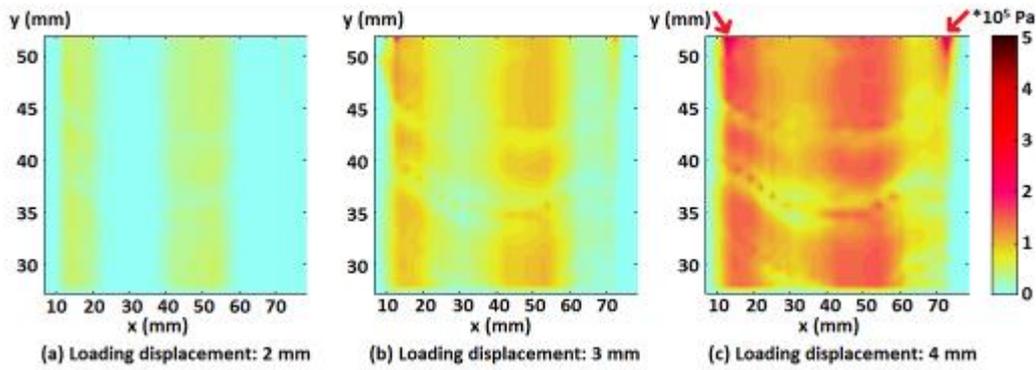

**Fig. 12** Stress distribution maps in the y direction of simulation results in Fig. 10.

According to the analysis of the two evaluation parameters: the residual wrinkles and the stress distribution, generally the greater the loading displacement, the less residual wrinkles, but the greater stress concentration at the two ends of the fix lateral sides, and at the position, the skin is easily injured. To better balance the two conflict parameters, the loading displacement $L_d = 3$ mm seems achieving the best performance.

*4.2 Optimizing $d_{fs}$*

Like the optimization of $L_d$, to optimize $d_{fs}$, three cases with $d_{fs}$ = 5.69 mm, 7.58 mm, 15.16 mm and constant $L_d = 3$ mm are studied, and the contour of residual wrinkles' depth for the three cases are shown in Fig. 13. As shown in Fig. 13, it is obvious that when $d_{fs}$ is 15.16mm, the residual wrinkles in Fig. 13c are more serious than the other two. This is because the stretching line becomes short, and the wrinkles locating outside the area $R_a$ are not enough stretched.

Meanwhile, the quantified parameters of the residual wrinkles are listed in Table 2. It seems that the appropriate $d_{fs}$ is 5.69 mm, as all of the evaluation parameters reach their minimum values. However, although the case of $d_{fs}$ = 5.69mm has a better P80 than the case of $d_{fs}$ = 7.58 mm, their P60 and area are similar. This means that the improvement of the former case is not significant with respect to the latter's, as only the area of very deep wrinkles decreased.

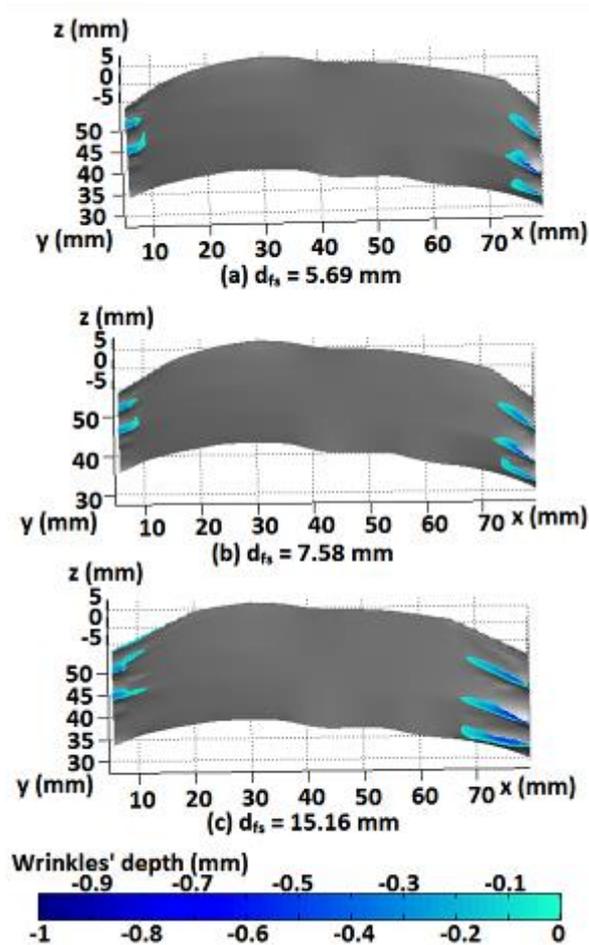

**Fig. 13** The stretching surgery simulation of forehead wrinkles in Fig. 1 under a constant $L_d$ of 3 mm and three different $d_{fs}$ of 5.69 mm, 7.58 mm and 15.16 mm.

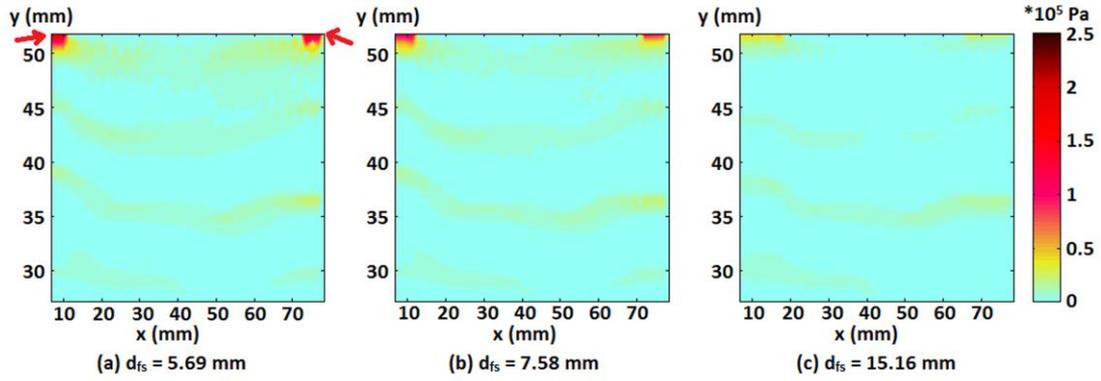

**Fig. 14** Stress distribution maps in the x direction of simulation results in Fig. 13.

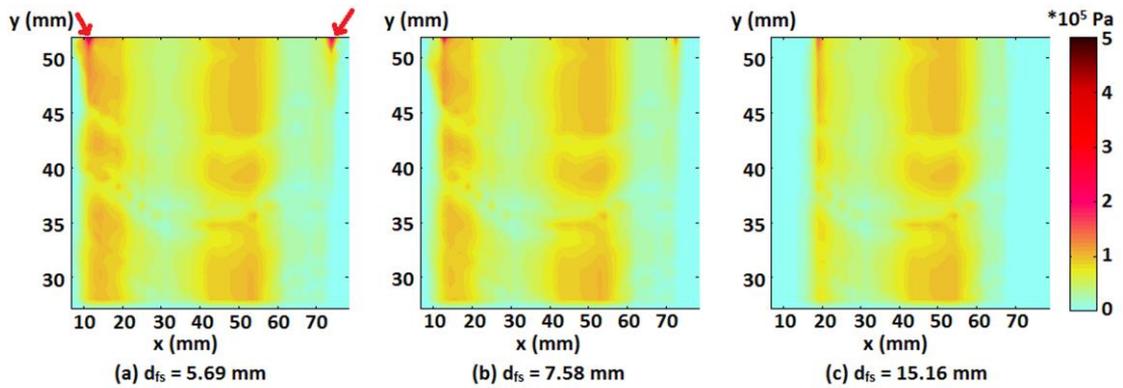

**Fig. 15** Stress distribution maps in the y direction of simulation results in Fig. 13.

**Table 2** The Quantified Parameters of the Residual Wrinkles of Simulation Results in Fig. 13

| No. | $L_d$/mm | $d_{fs}$/mm | Number of wrinkles | Max depth /mm | Mean depth /mm | P60 | P80 | Am/mm² | Area/mm² |
|---|---|---|---|---|---|---|---|---|---|
| 0 | 0 | 0 | 6 | 0.77 | 0.25 | 0.1132 | 0.0370 | 165.0290 | 390.6769 |
| 1 | 3 | 5.69 | 5 | 0.74 | 0.18 | 0.0454 | 0.0159 | 11.1343 | 47.7587 |
| 2 | 3 | 7.58 | 5 | 0.75 | 0.19 | 0.0466 | 0.0175 | 14.1482 | 53.6958 |
| 3 | 3 | 15.16 | 6 | 0.76 | 0.22 | 0.0724 | 0.0172 | 29.5608 | 94.9483 |

Corresponding to the Fig. 13, the stress distribution in the x and y directions are shown in Fig. 14 and Fig. 15, respectively. As shown in Fig. 14a, there exists a great stress in the regions marked by arrows and the stress of these regions decreases with the increase of $d_{fs}$, see Fig. 14b and 14c. This is because that the ends of the stretching line gradually get far away from the fixed lateral sides. When $d_{fs}$ is 5.69 mm, the stress exceeds $2.5 \times 10^5$ Pa, which indicates that the overstretching occurs in these regions[15].

On the contrary, when $d_{fs}$ is 15.16 mm, there are not any stress concentration and the maximum stress is $0.6 \times 10^5$ Pa. From Fig. 15, the stress in the y direction locating outside the area $R_a$ in Fig. 9 decreases and gradually disappears with the increase of $d_{fs}$. On the contrary, the stress in the $R_a$ region changes a little for all the three cases. Especially, when $d_{fs}$ is 5.69 mm, although the forehead skin is not overstretched in the y direction, as the maximum stress is $2.9 \times 10^5$ Pa, the stress concentration regions marked by the arrows in Fig. 15a increase dramatically[15].

According to the analysis of the residual wrinkles and the stress distribution of the three cases presented in Fig. 13. On the one hand, when $d_{fs}$ is 15.16 mm, although the stress is the lowest, the forehead wrinkles are stretched insufficiently. On the other hand, when $d_{fs}$ is 5.69 mm, forehead wrinkles are somewhat overstretched. Therefore, the appropriate $d_{fs}$ is 7.58 mm.

Table 3 Evaluation Conclusions of Simulation Results

| Evaluation $d_{fs}$/mm \ $L_d$/mm | 2 | 3 | 4 |
|---|---|---|---|
| 5.69 | Not enough stretched [a] | Overstretched | Overstretched |
| 7.58 | Not enough stretched | The appropriate parameters | Overstretched |
| 15.16 | Not enough stretched | Not enough stretched | Not enough stretched |

[a] "Not enough stretched" means that the wrinkles are not enough stretched and the area of residual wrinkles is obviously larger than that of the appropriate parameters.

In all, in Sections 4.1 and 4.2, we have optimized the two parameters $L_d$ and $d_{fs}$, and Table 3 concludes the evaluation by comprehensively considering the residual wrinkles and the stress distribution. According to Table 3, to optimize the surgical effect, the forehead wrinkles in Fig. 1 should be stretched with the loading displacement of 3 mm and the distance between the ends of stretching line and the fixed sides of 7.58 mm.

*4.3 Verification*

In order to verify the algorithm, the optimal surgical parameters are determined for forehead wrinkles of four volunteers in Table 4. Table 4 shows that both the maximum stresses of the four simulations in the x and y directions are less than the critical values beyond which the skin injures[15]. Moreover, since the initial wrinkles in Figs. 16a1-d1 are lower than that in Fig. 1, the smaller loading displacements $L_d$ and longer stretching lines are needed for satisfactory results. Meanwhile, under the optimal parameters, the simulation results of the initial forehead wrinkles Figs. 16a1-d1 corresponds to Figs. 16a2-d2, respectively. From Fig. 16, most of the wrinkles are stretched sufficiently under their optimal surgical parameters, and few residual wrinkles remain near the fixed sides, which are denoted by arrows. These indicate that the simulation results validate the present method, and the method can be used to evaluate the surgical effect of stretching forehead wrinkles'.

**Table 4** The Quantified Parameters of Original Wrinkles and Simulation Results in Fig. 16

| No. | Wrinkles | $L_d$/mm | $d_{fs}$/mm | Residual wrinkles' area/mm² | Max stress in x-direction /×10⁵Pa | Max stress in y-direction /×10⁵Pa |
|---|---|---|---|---|---|---|
| 1 | (a1) (a2) | 3 | 8.12 | 58.9368 | 1.1142 | 2.4193 |
| 2 | (b1) (b2) | 1 | 6.50 | 17.8281 | 0.2304 | 0.4433 |
| 3 | (c1) (c2) | 1.5 | 6.84 | 29.0873 | 0.4641 | 0.7080 |
| 4 | (d1) (d2) | 0.5 | 6.09 | 0 | 0.1332 | 0.3048 |

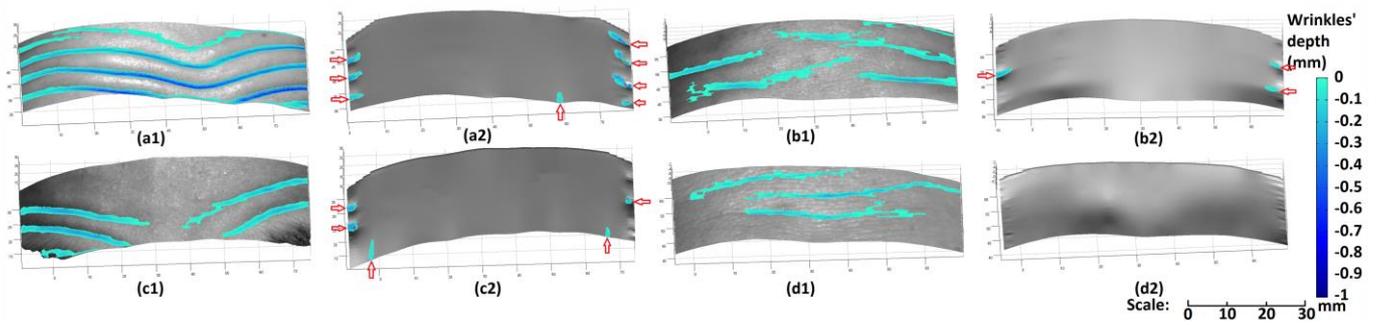

**Fig. 16** The original and stretched forehead wrinkles of different people.

## 5. CONCLUSIONS

In this paper, we have developed a finite element method to simulate the stretching process of the forehead wrinkles in rhytidectomy and determined the optimal surgical parameters. This study can be used to evaluate the surgical effect and optimize the surgical parameters.

The coronal-incision method in rhytidectomy under different parameters was simulated, and the preferable values of the simulated data were obtained through the analysis of the residual wrinkles and the stress distribution. According to the simulation, the greater loading displacement and the narrower distance between the stretching line and the fixed sides resulted in a greater stress concentration which could injure or fracture skin fibers. Thus it is crucial to choose the preferable values.

It is noted that the real forehead wrinkle stretching surgery in clinical rhytidectomy is more complicated than the simulation. Future work will include the simulation under more parameters or different surgical strategies such as the endoscopically assisted surgery[24-26].

## Appendix A. List of symbols

**Table A1** List of Symbols

| Symbol | Quantity | Symbol | Quantity |
|---|---|---|---|

| Symbol | Description | Symbol | Description |
|---|---|---|---|
| $\mathbf{Z_{sup}}$ | a three-dimensional rigid supporting contour | S | the projecting area of the square element |
| $\mathbf{Z_c}$ | the forehead wrinkles | $s_P$ | the trajectory of the particle P |
| $\varepsilon$ | the surface with wrinkles | $\mathbf{b_t}$ | the position of particle P at time t |
| $Z_f(x,y)$ | the fitting surface | $a_t$ | the acceleration vector of particle P at time t |
| $C_j$ | the control points | $v_t$ | the speed vector of particle P at time t |
| $N_{j,p}$ | the fitting polynomial | $\mathbf{b_{t+\Delta t}^{virt}}$ | the virtual position of particle P at time t+Δt if $\mathbf{Z_{sup}}$ does not exist |
| $n_x$ | the numbers of nodes along the x axis | $\mathbf{v_{t+\Delta t}^{virt}}$ | the virtual speed of particle P at time t+Δt if $\mathbf{Z_{sup}}$ does not exis |
| $n_y$ | the numbers of nodes along the y axis | O | the point that point P collides with the supporting contour |
| p | the order of the fitting polynomial | $\mathbf{b_{t+\Delta t}}$ | the position of particle P at time t+Δt |
| C | the control point matrix | $\mathbf{v_{t+\Delta t}}$ | the speed vector of particle P at time t+Δt |
| B | the approximate value of the second partial derivatives | Q | an arbitrary point n the supporting contour |
| $\lambda$ | the weight of the sum of the second partial derivatives | $w_z$ | the distance from the point Q to the position $\mathbf{b_{t+\Delta t}^{virt}}$ |
| $\mathbf{Z_{i-1}}$ | the result of the (i-1)th iteration | n | the number of the time step |
| $\mathbf{Z_d}$ | a surface that do not contain wrinkles | $f_{ad}$ | a damping force |
| P | an arbitrary particle of the forehead wrinkle | α | the damping coefficient |
| $m_p$ | the mass of the particle | $x_{P,0}$ | the initial position of the particle P at time 0 |
| t | time | $v_{P,0}$ | the initial speed of the particle P at time 0 |
| Δt | time step | $A_x$ | the cross-sectional areas in the x direction |
| $F_{P,t}$ | the tensile force vector applied on particle P at time t | $A_y$ | the cross-sectional areas in the y direction |
| $f_{P,t}^{int}$ | the internal force | $U_{x,0}$ | the initial length of skin element in the x direction |
| $f_{P,t}^{ext}$ | the external force | $U_{y,0}$ | the initial length of skin element in the y direction |
| E | the elastic modulus of skin | $t_0 = 0$ | the initial time |
| $f_{P,t}^{i}$ | the internal force exerted by the ith surrounding structure unit at time t | u | a unit normal vector of the point O on the supporting surface |
| $U_{i,0}$ | the initial length vectors of the ith tensile structure unit at time t | $L_d$ | loading displacement |
| $\Delta U_{i,t}$ | the elongation vectors of the ith tensile structure unit at time t | $d_{fs}$ | distance between stretching line and the fixed sides |
| Am | the area of wrinkles whose depth is greater than the mean depth of the initially unstretched forehead wrinkles | P60 | the area fractions of wrinkles, whose depth is over 60% of the maximum depth of the initially unstretched forehead wrinkles |
| $R_a$ | a region in Fig. 9, the stress in which changes a little under a constant $L_d$ and different $d_{fs}$ | P80 | the area fractions of wrinkles, whose depth is over 80% of the maximum depth of the initially unstretched forehead wrinkle |

*Disclosures*

The authors declare that they have no conflict of interest.

*Acknowledgments*

The authors thank Mr. Xinran Liu in University of Waterloo, Mr. Junjie Yuan, Professor Suiren Wan, Professor Yu Sun, Mr. Tongjing Zhu, Mr. Guanyu Zhao and Miss Jing Jin in Southeast University for their admirable work in establishing the measurement system, the evaluation system of residual wrinkles and their volunteer work in data acquisition.

This work was partially supported by the National Key R&D Program of China (No. 2017YFC0112801), the National Natural Science Foundation of China (Nos. 61127002, 31300780), the Nature Science Foundation of Suzhou (No. SYG201313), the Fundamental Research Funds for the Central Universities (No. 2242016R30014) and the Key Program of Shanghai Zhangjiang Special Development Fund (No. 1701-JD-D1112-030).

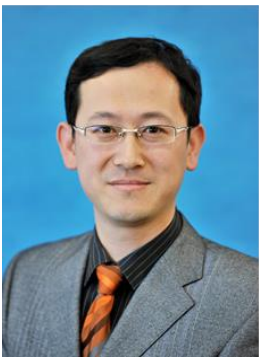

**Ping Zhou** received the B.S. degree in electronic engineering from University of Science and Technology of China (USTC) in 2002 and Ph.D. degree in biomedical engineering from USTC in 2007. From 2007, he was one of teachers in school of biological sciences and medical engineering, Southeast University, China. Now, he is an Associate Professor in the same school. His research interest includes stereo vision and image processing.

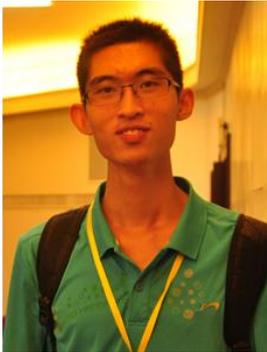

**Shuo Huang** was born in Yanzhou City, Shandong Province, China, in 1994. He got his B.S. and M.S. degree in biomedical engineering from Southeast University, Nanjing, China, in 2015 and 2018, respectively. From 2016 to 2017, he studied in the Joint Program between Southeast University and University of Rennes 1, and got a M.S. degree in electronic engineering from University of Rennes 1, Rennes, France. Since 2018, he has been working as an algorithm engineer in Shanghai United-imaging Healthcare Co., Ltd, Shanghai, China.

His research interests are numerical analysis and image processing.

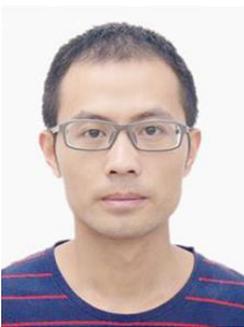

**Qiang Chen** was born in Changge City, Henan Province, China, in 1982. He got his B.S. degree from Xi'an University of Architecture & Technology in 2005 and M.S. degree from Southeast University in 2008 in Civil Engineering, respectively. In 2012, he received his doctor degree in Mechanics in Politecnico di Torino in Italy.

From 2012 to 2014, he was a Lecturer in the School of Biological Science & Medical Engineering. In 2014, he was promoted to be an Associate Professor in biomechanics. Dr. Chen's research interest is biomechanics.

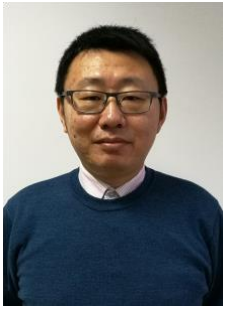

**Siyuan He** received the B.S. degree from the School of Materials Science and Engineering, Southeast University, China, in 1996, and the Ph.D. degree in the lab of Mechanical Systems and Concurrent Engineering (LASMIS) from the University of Technology of Troyes (UTT), France in 2005. He is currently with the School of Biological Sciences and Medical Engineering, Southeast University, China.

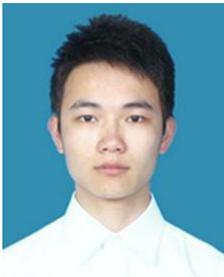

**Guochao Cai** got his B.S. and M.S. degree in biomedical engineering from Southeast University, Nanjing, China, in 2015 and 2018, respectively. Now, he is a software engineer in Natease Games, Hangzhou, China. He has done some research about the data structure, GPU acceleration and multithreading. Image processing and wrinkle reconstruction are also his research interests.

**Caption List**

**Fig. 1** A set of point cloud data of the selected forehead wrinkles.

**Fig. 2** The extraction of supporting contour. (a) The smooth forehead contour $\mathbf{Z_s}$. (b) Supporting contour $\mathbf{Z_{sup}}$ and the forehead wrinkles $\mathbf{Z_c}$.

**Fig. 3** The flowchart of supporting contour extraction.

**Fig. 4** The finite element meshing of the forehead wrinkles.

**Fig. 5** Discrete particle motion model, $s_P$ is the trajectory of the particle P.

**Fig. 6** The collision model.

**Fig. 7** The flowchart of the forehead wrinkle stretching method.

**Fig. 8** The extraction of the surface $\varepsilon$ with wrinkles. (a) binary image of wrinkles after preliminary extraction. (b) binary image of the noise-removed surface. (c) the final three dimensional extraction results.

**Fig. 9** Diagram of forehead wrinkles' stretching.

**Fig. 10** The stretching surgery simulations of forehead wrinkles in Fig. 1 under a constant $d_{fs}$ of 7.58 mm, and different $L_d$ of 2 mm, 3 mm and 4 mm.

**Fig. 11** Stress distribution maps in the x direction of simulation results in Fig. 10.

**Fig. 12** Stress distribution maps in the y direction of simulation results in Fig. 10.

**Fig. 13** The stretching surgery simulation of forehead wrinkles in Fig. 1 under a constant $L_d$ of 3 mm and three different $d_{fs}$ of 5.69 mm, 7.58 mm and 15.16 mm.

**Fig. 14** Stress distribution maps in the x direction of simulation results in Fig. 13.

**Fig. 15** Stress distribution maps in the y direction of simulation results in Fig. 13.

**Fig. 16** The original and stretched forehead wrinkles of different people.

**Table 1** The Quantified Parameters of the Residual Wrinkles of Simulation Results in Fig. 10

**Table 2** The Quantified Parameters of the Residual Wrinkles of Simulation Results in Fig. 13

**Table 3** Evaluation Conclusions of Simulation Results

**Table 4** The Quantified Parameters of Original Wrinkles and Simulation Results in Fig. 16

**Table A1** List of Symbols